# Upper critical field and fluctuation conductivity in the critical regime of doped SmFeAsO

I.Pallecchi [a], C.Fanciulli [a], M.Tropeano [a], A.Palenzona [b], M.Ferretti [b], A.Malagoli [a], A.Martinelli [b], I.Sheikin [c], M.Putti [a,d], C.Ferdeghini [a]

[a] *CNR-INFM-LAMIA and Università di Genova, via Dodecaneso 33, 16146 Genova, Italy*
[b] *DCCI, Università di Genova and CNR-IMEM unità di Genova, via Dodecaneso 31, 16146 Genova, Italy*
[c] *GHMFL, MPI-FKF/CNRS, 28 Avenue des Martyrs, BP 166, 38042 Grenoble, France*
[d] *ASC, NHMFL, FSU, 2031 E. Paul Dirac Dr., Tallahassee, FL 3231, USA*

**Abstract**

We measure magnetotransport of F doped SmFeAsO samples up to 28T and we extract the upper critical fields, using different criteria. In order to circumvent the problem of criterion-dependence $H_{c2}$ values, we suggest a thermodynamic estimation of the upper critical field slope $dH_{c2}/dT$ based on the analysis of conductivity fluctuations in the critical regime. A high field slope as large as -12T/K is thus extracted for the optimally doped sample. We find evidence of a two-dimensional lowest Landau level (LLL) scaling for applied fields larger than $\mu_0 H_{LLL} \sim 8T$. Finally, we estimate the coherence length values and we observe that they progressively increase with decreasing $T_c$. In all cases, the coherence length values along the c axis are smaller than the interplanar distance, confirming the two-dimensional nature of superconductivity in this compound.

**Introduction**

Since the recent discovery of superconductivity with $T_c$=26K in doped LaFeAsO [1], the class of these iron-based oxypnictides has been object of intensive research, which have lead to the current record of $T_c$=56K [2]. In compounds with general chemical formula REFeAsO (RE=La, Sm, Nd, Ce…), it is believed that doped REO layers act as charge reservoirs for high mobility FeAs planes, making oxypnictides very similar to the layered structure of high-$T_c$ cuprates. Beyond the theoretical challenge of understanding the unconventional pairing mechanism responsible for superconductivity, many studies have been devoted to superconductive properties of applicative appeal, such as the huge values of the upper critical field and its dependence on the particular rare earth RE [3]. Slopes of $H_{c2}(T)$ as large as about -9.3T/K have been measured in compounds with Sm. Such high $H_{c2}(T)$ implies a short coherence length that, together with the high $T_c$, the layered structure and the rather anisotropic Fermi surface, make these compounds candidate to present enhanced thermal fluctuations. In the case of the RE=Nd compound, coherence lengths along the c axis, $\xi_c$ varying from 0.26 to 0.9 nm have been evaluated [4,5,6] which are lower or comparable to the distance between the FeAs-layers. The occurrence of remarkable thermal fluctuations, indeed, has been observed by specific heat analysis.

The investigation of thermal fluctuation effects may offer several hints for the comprehension of these compounds. The identification of three-dimensional (3D) rather than two-dimensional (2D) thermal fluctuations is a crucial point, because the latter could imply the occurrence of pronounced dissipation in the mixed state, detrimental for applications, as observed in cuprates.

Among the oxypnictides, SmFeAs($O_{1-x}F_x$) presents high $T_c$ (~54 K [3,7,8]), the largest slope of $H_{c2}(T)$ and high anisotropy [9,10] which make this compound the most suitable for investigating fluctuation effects. In turns, however, due to the expected fluctuation contribution just above $T_c$, there is not univocal consensus about the criterion for the extraction of the upper critical field from transport experimental data. Alternatively, a thermodynamic determination of the transition temperature at different magnetic fields has been proposed which is thought to reflect an intrinsic behavior of the material; in this analysis, lower $H_{c2}$ slopes and consequently larger coherence lengths have been evaluated.

The same problem has been met with cuprates and the analyses of fluctuation effects of conductivity, specific heat, magnetization, Ettinghausen, Hall and Nernst effects have been suggested to provide a reliable tool for this aim, provided that fluctuation effects are significant in an appreciable temperature range around $T_c(H)$. Fluctuation effects are quantified by the Ginzburg number, which can be expressed for a 3D superconductor by $Gi_{3D}=(\pi\kappa^2\xi_0 K_B T_c \mu_0/2\Phi_0^2)^2$ and, in presence of a magnetic field, by its magnetic field dependent counterpart [11] $Gi_{3D}(H)=(2H\cdot Gi_{3D}^2/H_{c2})^{2/3}$, where $\kappa=\lambda_0/\xi_0$ is the Ginzburg-Landau parameter, $\lambda_0$ is the London penetration depth, $\xi_0$ is the coherence length, $H_{c2}$ is the zero temperature limit of the upper critical field, $K_B$ is the Boltzmann constant, H the applied magnetic field and $\Phi_0$ is the flux quantum. In conventional superconductors $Gi_{3D}\sim10^{-5}$, while in high-$T_c$ cuprates it is ~$10^{-3}$-$10^{-2}$, making fluctuations significant in an experimentally accessible temperature window ~$Gi_{3D}\cdot T_c$; experimental results indicate that the actual temperature window of fluctuations is even larger than this estimate. In the RE=Sm compound, assuming $T_c\approx54K$, $\lambda\approx190$ nm [12,13] and $\xi_c\approx0.6$ nm, which is an average value for the RE=Nd compound, we estimate $Gi_{3D}\sim8\cdot10^{-3}$, so that a similar situation to the case of high-$T_c$ cuprates occurs. For a 2D superconductor, that could be more appropriate for the case of SmFeAs($O_{1-x}F_x$), the Ginzburg number can be expressed as $Gi_{2D}=K_B T_C/E_F$ [14]. From Hall effect and effective mass data of ref. , this yields approximately $Gi_{2D}\approx0.01$, a remarkably high value.

In this paper, we measure magnetotransport SmFeAs($O_{1-x}F_x$) samples with different doping x, in magnetic fields up to 28 T. From the analysis of conductivity fluctuations in high fields, in the critical regime close to and below the transition temperature, we find evidence of 2D behavior and we extract the high field linear slope of the upper critical field $H_{c2}(T)$, which is as large as

$dH_{c2}/dT \approx -12T/K$ in the most doped sample and progressively decreases with decreasing $T_c$. Consequently the coherence lengths values increase, still remaining smaller than the interplanar distance, which confirms the 2D nature of superconductivity in these compounds.

**Fluctuation conductivity in the critical regime**

The classical theory of fluctuations near $T_c$ (hereafter, throughout the treatment of fluctuation effects, the transition temperature $T_c$ must be intended as the mean field transition temperature) predicts that the fluctuation conductivity $\Delta\sigma$ due to finite Cooper pair formation above $T_c$ is described in the 3D and 2D cases by [15,16] $\Delta\sigma_{3D} \propto \varepsilon^{-1/2}$ and $\Delta\sigma_{2D} \propto \varepsilon^{-1}$ laws, respectively, where $\varepsilon = \ln(T/T_c) \approx (T-T_c)/T_c$. In the Lawrence-Doniach picture for layered systems [17], a crossover from 2D to 3D behavior is expected to occur approaching $T_c$, due to the divergence of the coherence length. Within this Gaussian approximation, the fluctuation conductivity is therefore predicted to diverge close to the transition temperature. This nonanalyticy is removed if interactions between fluctuations are taken into consideration, which is necessary in the so called critical regime, that is in a region of the $T_c$-H phase diagram close to $T_c(H)$. The temperature range in which critical fluctuations should be observed is proportional to $\xi_0^{-6}$ in the 3D case, that is an experimentally accessible range of the order of 1K for high-$T_c$ superconductors [18]. In a magnetic field this range is expected to increase further as $\sim H^{2/3}$ or $\sim H^{1/2}$ in 3D and 2D, respectively [19].

In the critical regime, in the limit of strong magnetic fields, a scaling form for thermodynamic functions is expected. If quasiparticles are confined in the lowest Landau level (LLL), transport becomes of one-dimensional (1D) character along the direction of the applied field. Fluctuation effects close to the superconducting transition are further enhanced by the lower effective dimensionality of the system. Ullah and Dorsey [20,21] calculated the fluctuation conductivity $\Delta\sigma$ including the free energy quartic term within the Hartree approximation and obtained a scaling law for $\Delta\sigma$ in magnetic fields, in terms of unspecified scaling functions $F_{2D}$ and $F_{3D}$, valid for 2D and 3D superconductors, respectively:

$$\Delta\sigma(H)_{2D} = \left(\frac{T}{H}\right)^{1/2} F_{2D}\left(A\frac{T-T_c(H)}{\sqrt{T \cdot H}}\right) \quad (1.a)$$

$$\Delta\sigma(H)_{3D} = \left(\frac{T^2}{H}\right)^{1/3} F_{3D}\left(B\frac{T-T_c(H)}{(T \cdot H)^{2/3}}\right) \quad (1.b)$$

where A and B are characteristic constants of the material. Such functional dependence still holds even if more than one (but only a few) higher Landau levels are involved, as long as the inter-LL interaction is negligible [22,23]. However the scaling cannot account for the shape of the resistive transition in case the broadening is due to the dissipative flux line motion rather than to fluctuation effects. The field $H_{LLL}$ above which the LLL approximation should hold fulfils the condition that the Landau level spacing is larger than the Landau level interaction energy. According to Tešanovič and coworkers [22], this condition is translated as $H_{LLL} \sim (Gi/16)(T/T_{c0})H_{c2}(0)$, with $T_{c0}=T_c(H=0)$. More practically, they suggest the thumb rule $H_{LLL} \sim H_{c2}/3$.-

The LLL scaling has been successfully applied to high-$T_c$ cuprates. Polycrystalline samples [27,24] display similar behavior as single crystals. A 3D behavior has been observed in the case of $YBa_2Cu_3O_x$ [25,26,24] and a 2D one in the case of $Tl_2Ba_2CaCu_2O_x$ [27] and $Bi_2Sr_2CaCu_2O_8$ [28]. $\mu_0 H_{LLL}$ values are found to range from 1T [25,27,29,24] to ~10T [30,26] for high-$T_c$ cuprates.

**Experimental magnetotransport results**

We prepare polycrystalline $SmFeAs(O_{1-x}F_x)$ samples with nominal doping x=0.07, 0.1 and 0.15 following the steps described in [31]: once SmAs is synthesized from pure elements, it is reacted with stoichiometric amounts of Fe, $Fe_2O_3$, and $FeF_2$ at high temperature. Single phase formation is checked in all the samples by X-rays analysis.

Resistivity measurements from 4K to 300K are carried out in magnetic fields up to 9 T in a PPMS Quantum Design system and up to 28 T at Grenoble High Magnetic Field Laboratory.

In the left-hand panel of figure 1, we present resistivity curves normalized at $\rho(300K)$ of the three $SmFeAs(O_{1-x}F_x)$ samples analyzed in this paper. It can be seen that the resistivity curves exhibit metallic behavior, with a tendency to saturate above ~160K. Below this temperature $\rho$ decreases linearly with temperature and its slope progressively decreases with decreasing $T_c$ of the sample, with a monotonic tendency to a weaker metallicity from the optimally doped to the underdoped sample.

The superconducting transitions are quite rounded, making it difficult to univocally determine the transition temperatures and width. In Table I the values of $T_c$ obtained by the criterion of 50% resistivity drop and of $\Delta T_c$ obtained as the temperature interval between 90% and 10% resistivity drops are listed. The x=0.15 sample presents $T_c$=51.5 K, the x=0.1 sample $T_c$=41.5 K and the least doped sample $T_c$=33 K.

In the three right-hand panels of figure 1, the zoomed transition regions in magnetic fields from zero to 28T for the three samples are shown. The typical fan-shaped broadenings of the transitions are well visible for all the samples, indicating an important contribution of fluctuations.

From these magnetotransport data, we extract the characteristic critical fields, using different criteria. The criterion proposed in ref. [32] consists in finding the intersection $\rho^*(T^*)$ between linear extrapolations of the $\rho(T)$ curves above and below $T_c$. Hence, the temperatures at which the resistivity is 90% and 10% of $\rho^*$ allow to trace $H^*_{90\%}(T)$ and $H^*_{10\%}(T)$, respectively. Another common criterion consists on extrapolating linearly the normal state resistivity, then trace the parallel straight lines at 90% and 10% of the normal state extrapolation and find the respective intercepts with the $\rho(T)$ curves; this procedure identifies $H_{90\%}(T)$ and $H_{10\%}(T)$, respectively. In figure 2, we plot all the above critical fields for the three samples. In Table II we report the values of the high field ($\geq 8T$) linear slopes of such critical fields. Those extracted from the tails are believed to represent the irreversibility fields, and have similar values for all the samples, around -2T/K÷-3T/K, depending on the used criterion. The critical fields extracted from the onsets of the transitions represent the upper critical fields $H_{c2}$. As transport always follows the less resistive path, in polycrystalline samples with randomly oriented grains, the measured upper critical fields are the largest ones, that is those of the grains whose c axis is perpendicular to the applied field (usually labeled $H_{c2\|ab}$). These upper critical field slopes vary from -5T/K÷-20T/K, depending on the criterion. However, by inspecting Table II it is clear that the variations due to different criteria are comparable to the variations from sample to sample. This sensitivity to the used criterion makes it also difficult to compare different values in literature. In the next section, we propose a criterion to extract a thermodynamic $H_{c2}$, that is based in conductivity fluctuations in the critical regime.

**Experimental fluctuation conductivity results**

We extract the fluctuation conductivity $\Delta\sigma$ as the difference between the normal state conductivity $\sigma_n$ and the measured conductivity $\sigma$. The former $\sigma_n$ is obtained from extrapolation of the resistivity measured in a temperature interval suitably chosen for each sample, roughly 30K around $2T_c$, where fluctuation contribution is assumed to be negligible. For the three samples, $\Delta\sigma$ in zero field as a function of $\varepsilon$ are plotted in the insets of figure 3; therein, the solid lines indicate the 2D asymptotic behavior $\Delta\sigma_{2D}\propto\varepsilon^{-1}$. In the Gaussian regime, not too close to the transition (0.01<$\varepsilon$<0.1), $\Delta\sigma$ exhibits a well defined asymptotic 2D behavior $\propto\varepsilon^{-1}$, delimited at either sides by sharp crossovers to the critical regime for $\varepsilon$<0.01 and the high-temperature short-wavelength regime for $\varepsilon$>0.1.[33] It can be also seen that the 2D behavior $\Delta\sigma_{2D} \propto \varepsilon^{-1}$ is extended up to $\varepsilon$~0.1, which means 3 to 5K above $T_c$ (it is worth noting that $T_c$ values extracted by this fluctuation analysis fall very close to the 50% resistivity drop values reported in Table I). This interval is larger than the transition widths and the presence of some inhomogeneity, which may yield a distribution of $T_c$, does not mask the characteristic behavior of Gaussian fluctuations. However, to avoid possible ambiguities yielded by any $T_c$ distribution, in the following of the paper we focus on the shape of the critical fluctuation conductivity at high fields; in this regime, the LLL scaling is not affected by a relatively small $T_c$ distribution, but rather accounts for the in-field broadening of the transition, which is huge in our samples. For example, in the sample S3, $\Delta T_c(H=0)$~2K becomes as large as 7K at 8T and 17K at 28T.

In order to verify the 2D nature of fluctuations, in the main panels of figure 3, we plot $\Delta\sigma \cdot (H/T)^{1/2}$ versus $(T-T_c(H))/(T \cdot H)^{1/2}$, expected for the 2D LLL scaling, for the three samples. It is apparent that the 2D scaling succeeds in making the curves collapse in an extended temperature range and for fields above $\mu_0 H_{LLL}$~8T. To compare the 2D with the 3D scaling in figure 4, according to equations (1.a) and (1.b), we plot the $\Delta\sigma \cdot (H/T)^{1/2}$ versus $(T-T_c(H))/(T \cdot H)^{1/2}$, expected for the 2D LLL case (left-hand panel) and $\Delta\sigma \cdot H^{1/3}/T^{2/3}$ versus $(T-T_c(H))/(T \cdot H)^{2/3}$ expected for the 3D LLL case (right-hand panel) for the one of the samples as representative of all, S2, in linear (main panels) and semilogarithmic (insets) scales. The maximum and minimum values of the horizontal axes for the 2D and 3D cases are chosen so that they correspond to the same temperature interval. The 2D relationship provides a better scaling of data both below and above $T_c$; indeed, in the 3D case, the curves at different fields open like a fan above $T_c$ and tend to depart from each other also below $T_c$. On the contrary, in the 2D case, the curves above 6T overlap almost in the whole range, being the broadening simply due to random scattering of experimental data. 2D character means that the relevant coherence length $\xi_c$ is smaller than the inter-plane spacing, or, equivalently, that the Josephson coupling between adjacent superconducting planes is smaller than intraplane condensation energy. The success of the scaling also indicates that the fan-shaped transitions displayed in figure 1 are primarily due to thermodynamic fluctuations of the superconducting order parameter, rather than to dissipative motion of flux lines. The latter effect may indeed come into play at low negative values of the variable in the horizontal axes $(T-T_c(H))/(T \cdot H)^{1/2}$, where the fluctuation conductivity curves start to depart from one another. This occurs several degrees K below $T_c$, just at the tail of the transitions.

The crossover value $\mu_0 H_{LLL}$~8T is worth some considerations. In $YBa_2Cu_3O_x$, lower values are generally found.[25,24,34] The criterion proposed by Tešanovič[22] $H_{LLL}$~$H_{c2}/3$ suggests that the higher upper critical fields of $SmFeAsO_{1-x}F_{1-x}$ as compared to those of $YBa_2Cu_3O_x$ may account for the difference in the $H_{LLL}$ values.

We can use the validity of the 2D LLL scaling to extract $T_c(H)$, treated as fitting parameter. The so obtained thermodynamic $H_{c2}$ curves are reported in Table II and shown as a function of temperature in figure 2, together with the $H_{c2}$ curves extracted from other criteria. Remarkable values of the thermodynamic $\mu_0 H_{c2}$ linear slope at high fields are found, namely -12T/K for S3, -7.5T/K for S2 and -7T/K for S1. The value of -12 T/K obtained in the most doped sample is the largest reported for oxypnicides. With such $dH_{c2}/dT$ value, the single band Werthamer-Helfand-Hohenberg (WHH) formula [35] yields in the zero temperature limit $\mu_0 H_{c2} \approx 0.693 T_c \mu_0 |dH_{c2}/dT|_{T_c} \approx 400T$. This huge upper critical field is well above the paramagnetic limit and suggests that a Pauli limiting behavior should be evident in these compounds at high fields [36].

With decreasing $T_c$ the slope values progressively decrease; this is in disagreement with was observed in LaFeAsOF where $H_{c2}$ evaluated with the 80% criterion has been found to follow a non monotonic trend with $T_c$.[37] In conventional superconductors such as A-15 and $MgB_2$,[38] $dH_{c2}/dT$ increases with decreasing $T_c$ because the $T_c$ reduction is often accompanied by an increase of disorder that carries the superconductor into the dirty limit. In short coherence length superconductors like high-$T_c$ cuprates, the dirty limit is not likely achieved and effects related to the doping prevail. This might be the case of oxypnictides, even if disorder effects on $H_{c2}$ have been suggested for LaFeAsO$_{0.9}$F$_{0.1}$ . However, lower H$_{c2}$ values of RE=La as respect to the RE=Sm compounds indicate that the latter should present lower intrinsic coherence length, more unlikely affected by disorder. If we neglect disorder effects, the simultaneous reduction of dH$_{c2}$/dT and T$_c$ can be explained by a reduction of the coupling that suppresses both the properties.

From H$_{c2}$ slope values, taking into account that in polycrystals H$_{c2}$ parallel to the ab-planes is actually probed and assuming an anisotropy factor γ=ξ$_{ab}$/ξ$_c$ ranging from 5 to 9 [9,10] the coherence lengths can be evaluated by means of the upper critical field obtained from the WHH formula. In Table III, the values for all the samples are calculated. Despite the large uncertainty on the anisotropy factor γ which includes also a possible dependence of γ on doping, we notice that the average values of ξ$_{ab}$ increase remarkably from ~23 Å to ~37 Å and those of ξ$_c$ increase from ~3.4 Å to 5.5 Å from sample S3 to S1. This suggests that moving from optimal to underdoped regimes a crossover from 2D to 3D regime could occur, even if in our case ξ$_c$ remains always smaller than the interplanar spacing s=8Å, confirming the 2D nature of superconductivity in the measured samples.

An opposite behavior was observed in YBa$_2$Cu$_3$O$_x$ where thermal fluctuations change from 2D to 3D nature moving from the underdoped to the optimally doped regime. [39,40] This opposite trend implies an important difference between oxypnictides and high-T$_c$ cuprates. In the latter ones, the doping changes substantially the transport properties from insulating to metallic and increases the coupling between the conducting planes. As a consequence, with increasing doping, the anisotropy progressively decreases and when its value is such that ξ$_c$=ξ$_{ab}$/γ becomes comparable to the interlayer spacing a crossover from 2D to 3D fluctuation regime occurs. Oxypnictides are metallic compounds: fluorine doping determines the occurrence of superconductivity by breaking symmetries of the Fermi surface, but it does not change substantially the transport properties [41] and anysotropy it is not expected to change substantially. Thus, a possible crossover from 2D to 3D fluctuation regime with doping could only be driven by the progressive increase of the coherence length with decreasing T$_c$.

**Conclusions**

We study fluctuation conductivity in F doped SmFeAsO polycrystalline samples, in the critical regime close and below T$_c$(H). At fields above μ$_0$H$_{LLL}$~8T, the conductivity curves obey the 2D LLL scaling, similarly to the BSSCO superconductor family. Such scaling allows to extract a thermodynamic estimation of the upper critical field, whose high field slope is as large as -12T/K in the most doped sample, in fair agreement with the values extracted with the criterion of 90% drop of the resistivity from its normal state value. The corresponding coherence lengths along the c axis turn out to be from 3Å to 6Å, always smaller than the interplanar spacing s=8Å, consistently with the 2D nature of superconductivity in these compounds.

**Acknowledgements**
This work is partially supported by Compagnia di S. Paolo and by the European Commission from the 6th framework programme "Transnational Access - Specific Support Action", contract N° RITA-CT-2003-505474 and by the Italian Foreign Affairs Ministry (MAE) - General Direction for the Cultural Promotion. The authors are grateful to A.Gurevich for helpful discussion.

**Figure and table captions**

**Table I**: Properties of the SmFeAs($O_{1-x}F_x$) samples: nominal doping x; transition temperature $T_c$, defined as the temperature at which the resistivity drops to 50% of its extrapolated normal state value and transition width $\Delta T_c$, defined as the temperature interval between 90% and 10% resistivity drops with respect to its extrapolated normal state value.

**Table II:** Linear slopes of the critical fields above 8T, extracted using different criteria, as explained in the text. In the last column are the slopes extracted from the LLL fitting of the fluctuation conductivity.

**Table III:** Values of the coherence length parallel and perpendicular to the superconducting planes $\xi_{ab}$ and $\xi_c$ in Å calculated from the WHH formula; the variability ranges are related to the range of literature values found for the anisotropy $\gamma=\xi_{ab}/\xi_c \approx 5 \div 9$.

**Figure 1:** (color online) Resistivity versus temperature curves of the three SmFeAs($O_{1-x}F_x$) samples in zero magnetic field (left-hand panel) and in magnetic fields up to 28T (right-hand panels). Due to an unknown geometric factor related to the sample density, resistivity absolute values suffer of an uncontrolled uncertainty and they have been normalized to their T=300K values.

**Figure 2:** (color online) Critical fields extracted from the in-field resistivity curves of the three samples, using different criteria (see text).

**Figure 3:** (color online) In the main panels, the plots of the quantity $\Delta\sigma \cdot (H/T)^{1/2}$ versus $(T-T_c(H))/(T \cdot H)^{1/2}$ for the three samples demonstrate the success of the 2D LLL scaling; in the insets, the curves of fluctuation conductivity $\Delta\sigma$ versus $\varepsilon=\ln(T/T_c)$ in the Gaussian regime and zero magnetic field are shown, together with the asymptotic 2D behavior $\Delta\sigma_{2D} \propto \varepsilon^{-1}$ indicated by continuous lines.

**Figure 4:** (color online) In the main panels, the plots of the quantity $\Delta\sigma \cdot (H/T)^{1/2}$ versus $(T-T_c(H))/(T \cdot H)^{1/2}$ (left-hand) and $\Delta\sigma \cdot (H/T^2)^{1/3}$ versus $(T-T_c(H))/(T \cdot H)^{2/3}$ (right-hand) expected for the 2D LLL and 3D LLL cases, respectively, are shown for the sample S2. In the insets, the same plots are displayed in semilogarithmic scale.

**Table I:**

| Sample | x | $T_c \pm \Delta T_c/2$ (K) |
|---|---|---|
| S1 | 0.07 | $33.0 \pm 1.5$ |
| S2 | 0.10 | $41.5 \pm 1$ |
| S3 | 0.15 | $51.5 \pm 1$ |

**Table II:**

| Sample | $\mu_0 dH^*_{10\%}/dT$ (T/K) | $\mu_0 dH_{10\%}/dT$ (T/K) | $\mu_0 dH^*_{90\%}/dT$ (T/K) | $\mu_0 dH_{90\%}/dT$ (T/K) | $\mu_0 dH_{flut}/dT$ (T/K) |
|---|---|---|---|---|---|
| S1 | -2.8 | -2.4 | -5.8 | -5.3 | -7.0±1 |
| S2 | -3.7 | -2.3 | -7.3 | -7.1 | -7.5±1 |
| S3 | -3.6 | -1.7 | -20.7 | -11.3 | -12±1 |

**Table III:**

| Sample | $\xi_{ab}$ (Å) | $\xi_c$ (Å) |
|---|---|---|
| S1 | 31÷42 | 6.3÷4.7 |
| S2 | 27÷37 | 5.4÷4.1 |
| S3 | 19÷26 | 3.9÷2.9 |

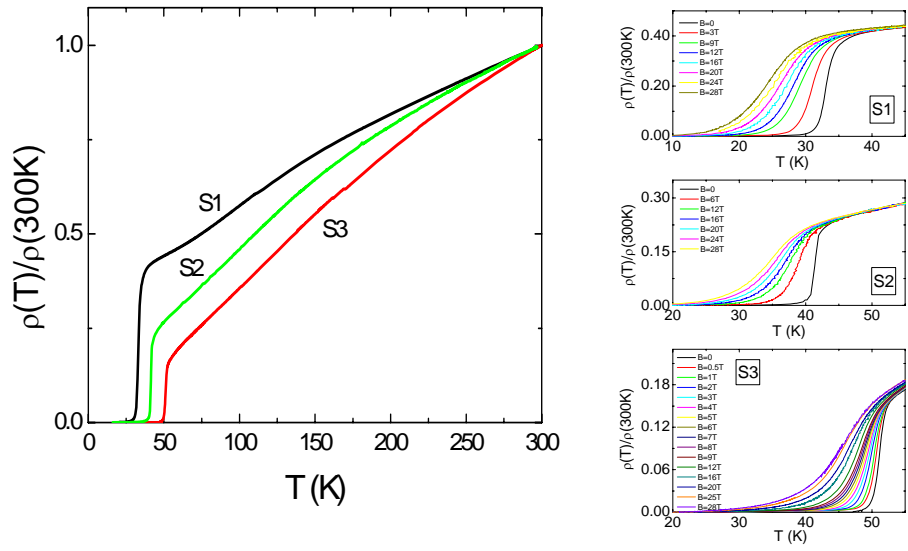

Figure 1

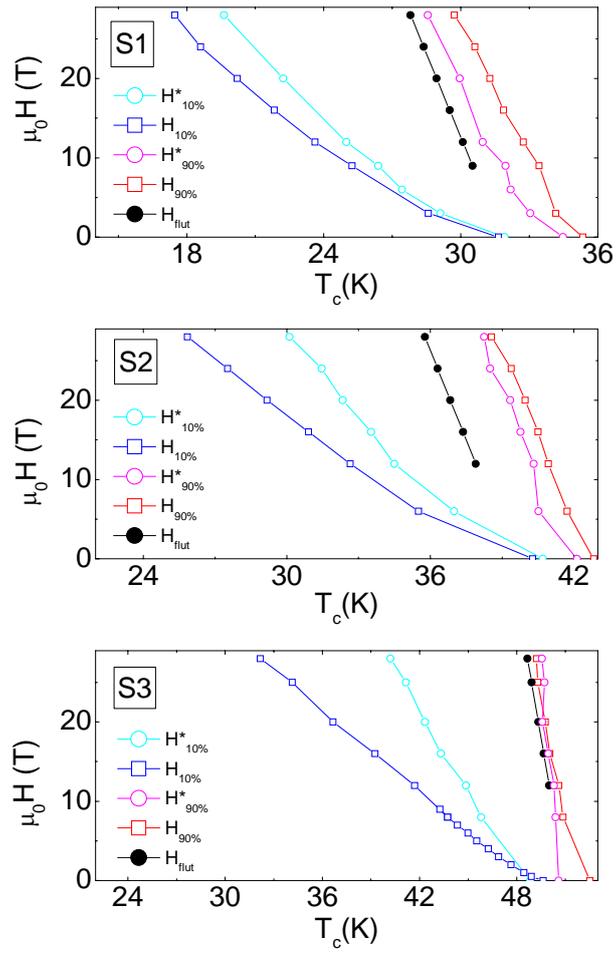

Figure 2

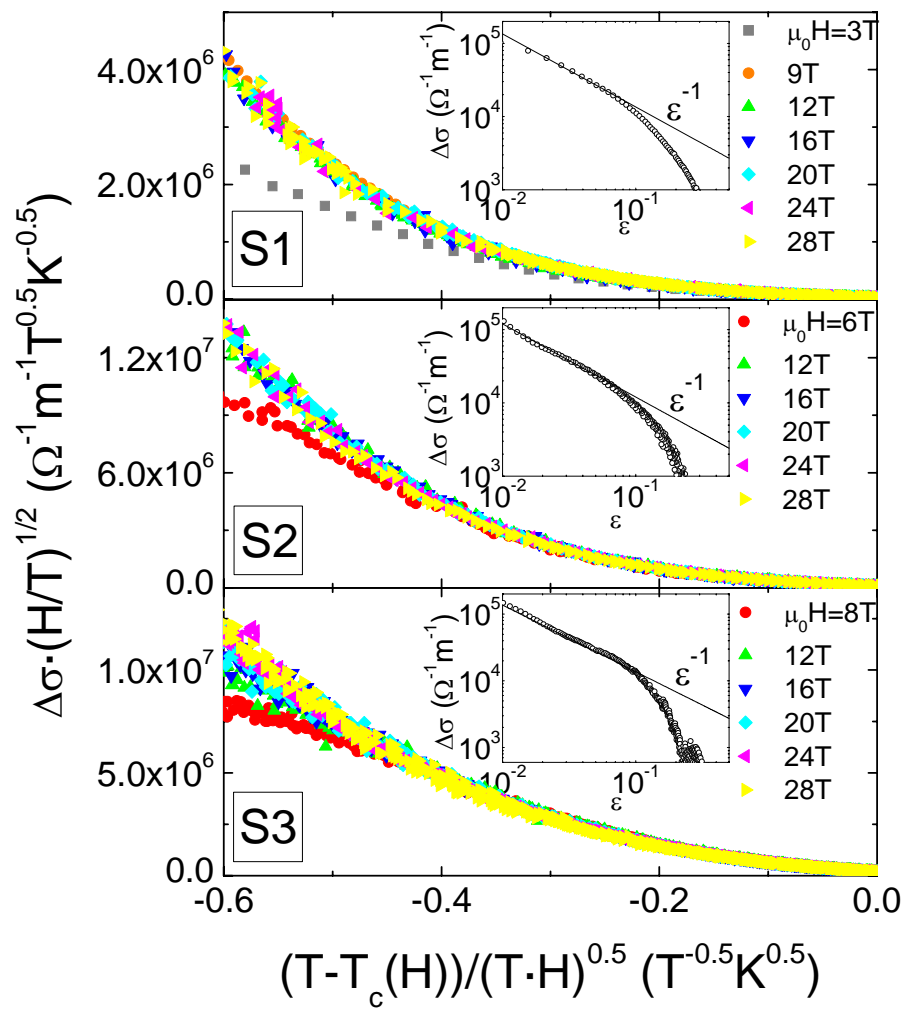

Figure 3

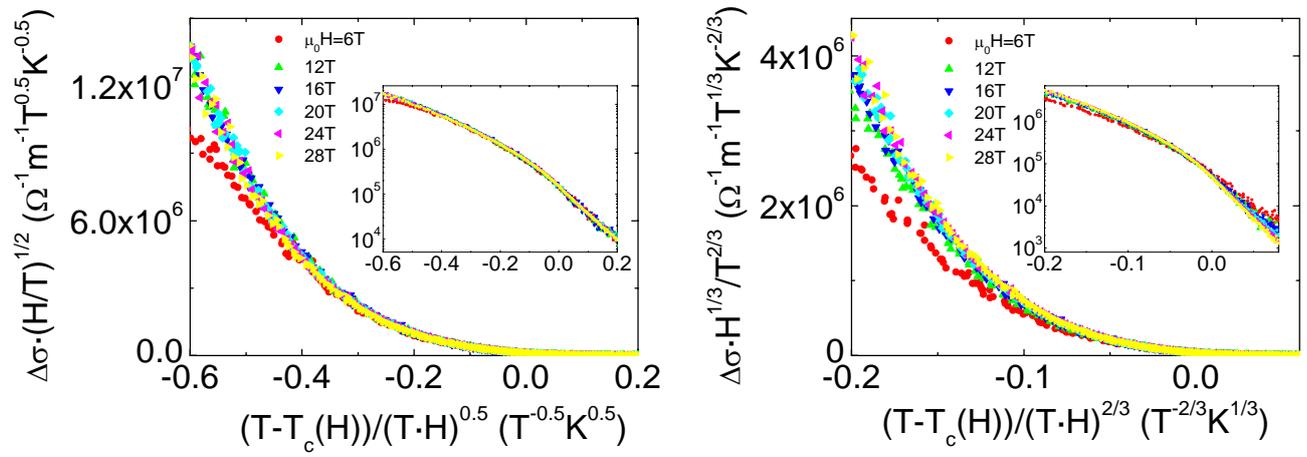

Figure 4